\documentclass[conference]{IEEEtran}
\IEEEoverridecommandlockouts

\usepackage{cite}
\usepackage{amsmath,amssymb,amsfonts}
\usepackage{algorithmic}
\usepackage{graphicx}
\usepackage{textcomp}
\usepackage[table]{xcolor}
\usepackage{booktabs}
\usepackage{hyperref}

\def\BibTeX{{\rm B\kern-.05em{\sc i\kern-.025em b}\kern-.08em
    T\kern-.1667em\lower.7ex\hbox{E}\kern-.125emX}}

\makeatletter

\def\ps@IEEEtitlepagestyle{%
  \def\@oddhead{}%
  \def\@evenhead{}%
  \def\@oddfoot{%
    \hfil
    \parbox[b]{0.96\textwidth}{%
      \centering
      \normalfont
      \fontsize{5.5pt}{6.3pt}\selectfont
      \textcopyright\ 2026 IEEE.
      Personal use of this material is permitted.
      Permission from IEEE must be obtained for all other uses,
      in any current or future media, including
      reprinting/republishing this material for advertising or
      promotional purposes, creating new collective works, for
      resale or redistribution to servers or lists, or reuse of
      any copyrighted component of this work in other works.
    }%
    \hfil
  }%
  \let\@evenfoot\@oddfoot
}

\makeatother

\begin{document}

\title{Alteron: A Tool for Behavioral Regression Testing Across NLP Classifier Versions}

\author{
\IEEEauthorblockN{Shazzad Hossain}
\IEEEauthorblockA{
University of Dhaka \\
bsse1203@iit.du.ac.bd
}
\and
\IEEEauthorblockN{Proma Chowdhury}
\IEEEauthorblockA{
University of Dhaka \\
bsse1132@iit.du.ac.bd
}
\and
\IEEEauthorblockN{Mridha Md. Nafis Fuad}
\IEEEauthorblockA{
University of Dhaka \\
fuad@iit.du.ac.bd
}
}

\maketitle
\thispagestyle{IEEEtitlepagestyle}

\begin{abstract}
Evaluating evolving Natural Language Processing (NLP) models is important for ensuring reliable behavior across updates, but standard benchmark metrics do not fully capture how model behavior changes across versions. Existing work has focused mainly on testing models in isolation rather than comparing successive versions in continuous integration workflows. We present Alteron, a tool for detecting behavioral regressions across NLP model versions with metamorphic testing. Alteron constructs a test corpus from labeled source examples and compares model versions on metamorphically transformed inputs. In an evaluation spanning 10 metamorphic relations (MRs), 4 model versions, and 3 model-update transitions, Alteron identified 16 behavioral regressions, 11 of which were release-blocking. The results show that common model updates can preserve overall task performance while still introducing undesirable behavior changes, and that behavioral checks across model versions can reveal failures that aggregate benchmark metrics alone do not capture. The tool is open-source and available at \href{https://github.com/shazzad5709/alteron}{https://github.com/shazzad5709/alteron}. A screencast demonstration is available at \href{https://youtu.be/szwiWW5O4do}{https://youtu.be/szwiWW5O4do}.

\end{abstract}


\section{Introduction}

Natural Language Processing (NLP) classifiers are widely used in modern software systems, making reliable evaluation essential as these models evolve over time. These models are frequently updated after deployment~\cite{renggli2019easemlci}. Before release, each updated version must be validated to ensure it behaves as intended. Standard evaluation methods typically rely on labeled benchmarks and aggregate metrics such as accuracy. These metrics summarize performance over a test set, but they do not capture how individual predictions change between model versions ~\cite{ribeiro2020checklist}. As a result, an updated model may perform better on some inputs while performing worse on others, even when aggregate metrics remain similar or improve. This can hide meaningful behavior changes on individual inputs.

Detecting behavior changes in NLP model updates is challenging as standard ML metrics summarize performance at the aggregate level and do not show how predictions shift on individual inputs. For example, a spam detection model may classify “Win a free iPhone now” as spam, but after an update misclassify “Win a free smartphone now,” although both messages express the same intent. We refer to this issue as behavioral regression. This motivates the need for evaluation methods that go beyond aggregate metrics and check whether updated models preserve consistent predictions on semantically equivalent inputs.

Recent work shows that Metamorphic Testing (MT) ~\cite{chen2018metamorphic} provides a practical way to expose behavioral failures in NLP systems without requiring labeled follow-up data~\cite{cho2025mtllmnlp}. Tool-oriented efforts such as LLMORPH~\cite{cho2025llmorph} and MDPMORPH~\cite{li2025mdpmorph} further demonstrate the practicality of automated MR-based testing for NLP and ML models. However, these approaches primarily focus on testing or analyzing a model in isolation. They do not address the continuous integration (CI) setting in which a new version must be compared against the previous accepted version to determine whether system evolution has introduced release-relevant behavioral regressions.

A\textsc{lteron} addresses this problem by treating metamorphic testing as a version-to-version behavioral regression check for NLP classifiers in continuous integration workflows. A\textsc{lteron} applies a predefined set of metamorphic relations to labeled source examples, stores the resulting source and follow-up pairs as a reusable test corpus, and records the predictions and MR outcomes for both the new version and the previous accepted version. It then compares the two versions only on source examples that both versions classify correctly. On that shared subset, A\textsc{lteron} measures how much the MR pass rate changes from one version to the next. The resulting reports distinguish blocking from non-blocking regressions and can be consumed directly in release pipelines through a machine-readable CI outcome.

We present a pilot evaluation of A\textsc{lteron} on 9 dataset profiles spanning sentiment analysis, natural language inference, and generic robustness settings. The pilot uses 4 model versions connected by retraining, distillation, and quantization transitions. The evaluation measures whether version-to-version decision behavior is preserved across controlled model updates over a fixed metamorphic test suite. The pilot identified 16 behavioral regressions, 11 of which were release-blocking, showing that A\textsc{lteron} can detect release-relevant regressions that are not apparent from source-side accuracy alone, particularly in generic robustness profiles. The paper contributes an operational tool based on metamorphic testing for NLP model evolution, a behavioral regression testing workflow for continuous integration, and a pilot demonstration of its use across realistic update scenarios.
\vspace{-2mm}
\section{Alteron}

\begin{figure}[t]
    \centering
    \includegraphics[width=0.9\columnwidth]{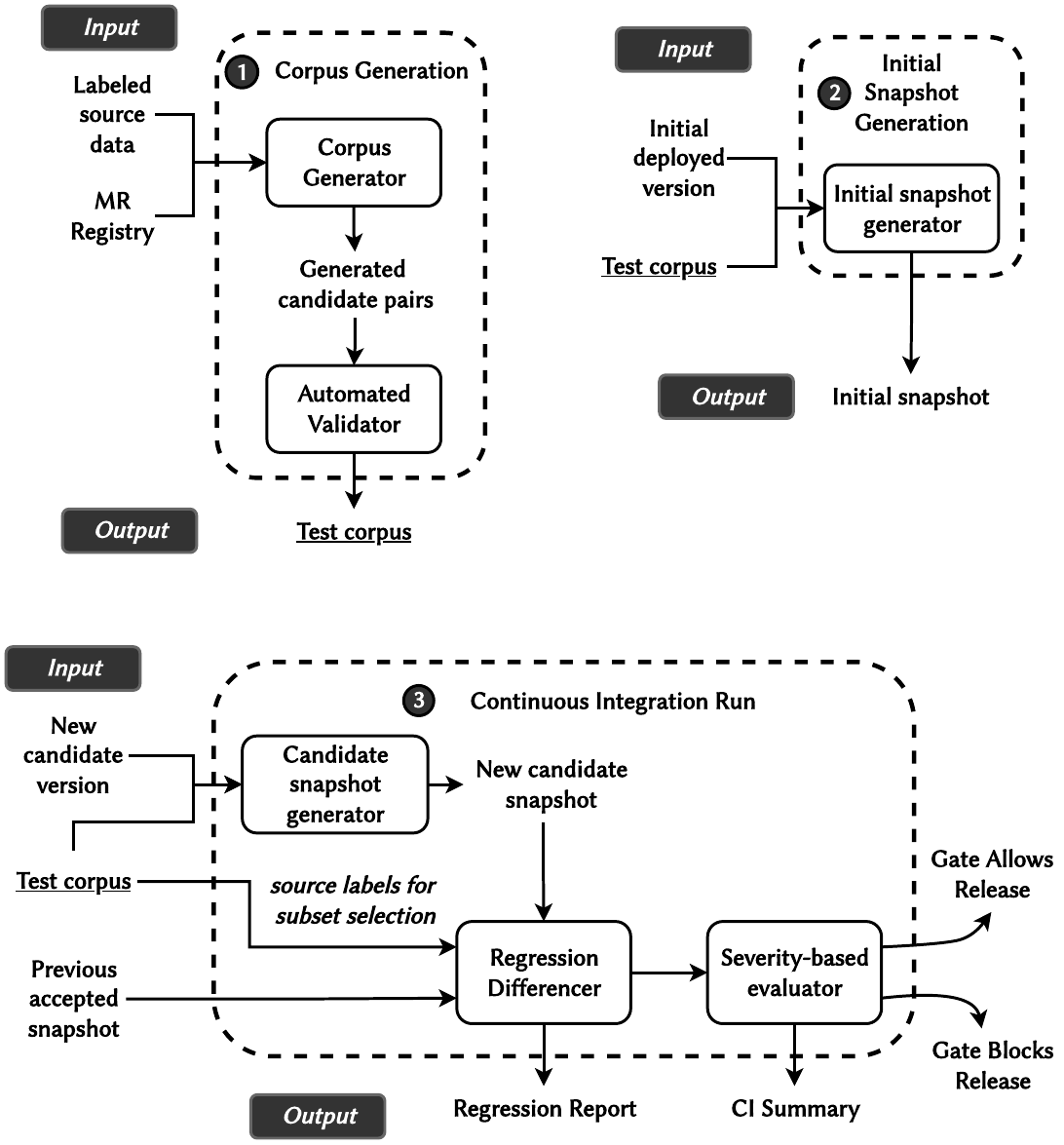}
    \caption{Logical workflow of \textsc{Alteron}.}
    \label{fig:architecture}
\end{figure}

A\textsc{lteron} is a behavioral regression testing tool for version-to-version evaluation of NLP classifiers in CI workflows. Figure~\ref{fig:architecture} illustrates the logical workflow of \textsc{Alteron}. Given a previously accepted model as the baseline, a new candidate model, a selected set of MRs, and a fixed test corpus, it produces behavioral snapshots, regression reports, and a machine-readable CI summary that indicates whether any detected regressions should block the candidate model from progressing through continuous integration.

\subsection{Input}

A\textsc{lteron} takes inputs across four categories:

\textbf{Model versions.} A\textsc{lteron} evaluates an NLP classifier through a framework-agnostic interface. Two versions of the model are required. The baseline represents the previously accepted model, and the candidate represents the new version under evaluation. The model must return a predicted label and a confidence score for each input.

\textbf{Labeled source data.} A\textsc{lteron} requires labeled source examples for the NLP task under test. Source labels are used to build the matched subset, which retains only examples that both versions classify correctly. They are also used by task-specific metamorphic relations to determine input eligibility.

\textbf{Metamorphic relations (MRs).} The user selects which MRs to apply from a registry. The registry draws on the MR catalog of Cho et al.\cite{cho2025mtllmnlp}, a systematic review of 191 MRs across 24 NLP tasks. A\textsc{lteron} currently implements ten MRs across three families. These families are summarized in Table~\ref{tab:mr-families}. The sentiment-specific family includes negation insertion, period-to-exclamation replacement, emphasizing adverb insertion, and noun/adjective uppercasing. The natural language inference (NLI)-specific family includes same-gender substitution, cross-gender substitution, and hypothesis negation. The generic robustness family includes space injection, capitalization change, and keyboard typo simulation.

\begin{table}[t]
\caption{MR families in A\textsc{lteron}.}
\vspace{-5mm}
\label{tab:mr-families}
\begin{center}
\footnotesize
\begin{tabular}{p{1.4cm} p{3.2cm} p{1.8cm} p{1.0cm}}
\toprule
\textbf{Family} & \textbf{Representative MRs} & \textbf{Expected Relation} & \textbf{Tasks} \\
\midrule
Sentiment & Negation insertion, adverb insertion, uppercasing & Flip / monotonic & SA \\
NLI & Gender substitution, hypothesis negation & Flip / invariant & NLI \\
Generic & Space injection, capitalization change, keyboard typo & Invariant & SA, NLI, topic \\
\bottomrule
\end{tabular}
\end{center}
\end{table}

\textbf{Baseline artifacts.} For CI regression checking, A\textsc{lteron} requires a fixed test corpus of validated MR test pairs and a behavioral snapshot of the previous accepted version over that corpus. The fixed test corpus is generated once and reused across model updates. The stored snapshot is reused until a new version is accepted, at which point the snapshot of the new version becomes the next baseline for comparison.

\subsection{Output}

A\textsc{lteron} produces four categories of output.

\textbf{Fixed test corpus.} A\textsc{lteron} produces per-MR CSV files that store validated source and follow-up pairs. It also creates a SHA-256 manifest, a rejection log explaining skipped inputs, and sample artifacts for manual review. The fixed test corpus is a verified set of source and follow-up pairs that is reused across model versions. This ensures that any observed differences come from changes in the model rather than the test data.

\textbf{Behavioral snapshots.} For each model version, A\textsc{lteron} records per-MR predictions on source and follow-up inputs, confidence scores, MR pass or fail outcomes, and fairness-regression flags where applicable.

\textbf{Regression reports.} For each MR, A\textsc{lteron} reports how model performance changes on the original input examples, how many matching cases were evaluated, and how the MR pass rate changes on that subset. It also indicates whether the observed change is considered a behavioral regression, the MR’s severity level in the registry, and whether the regression is severe enough to fail the CI check.

\textbf{CI result.} A\textsc{lteron} emits a machine-readable summary of the CI run together with an exit code. The summary records the overall outcome, including the selected profile, compared model versions, regression threshold, number of blocking regressions, and the generated report paths, while detailed MR-level reasons for failure are provided in the regression report. A\textsc{lteron} returns exit code \texttt{0} when no blocking regression is found and \texttt{1} otherwise, making it directly usable as a CI gate.

\subsection{Process}

A\textsc{lteron} operates in three stages:

\textbf{Corpus generation.} For each source input and selected MR, A\textsc{lteron} applies the transformation function to produce a follow-up input and then runs MR-specific automated checks to validate the pair. Invalid or ineligible pairs are recorded in a rejection log. Accepted pairs are written to a fixed test corpus with a SHA-256 hash manifest for integrity checks.

\textbf{Initial snapshot generation.} Each model version is evaluated on the fixed test corpus produced during corpus generation. For every source and follow-up pair, A\textsc{lteron} records the predicted labels, confidence scores, and whether the model’s outputs satisfy the expected metamorphic relation.

\textbf{CI run.} In each CI run, A\textsc{lteron} first evaluates the new candidate model on the fixed test corpus to generate a new snapshot. It then compares that snapshot against the stored snapshot of the previous accepted version through regression differencing. For each MR, this comparison uses only source examples that both versions classify correctly, so that the measured change reflects behavioral differences rather than regular classification error on the source inputs. A\textsc{lteron} flags a behavioral regression when the change in MR pass rate on that subset falls below a configurable threshold. Finally, A\textsc{lteron} evaluates the resulting regression signals under a configured CI profile to determine the reported outcome. A\textsc{lteron} provides built-in execution profiles to balance speed and coverage. The \texttt{pr-fast} profile samples a small number of examples per MR to provide rapid pull-request feedback, whereas the \texttt{release-full} profile evaluates the complete fixed test corpus for a full release-time check. Both profiles can be customized. If a blocking regression is detected under the active profile, A\textsc{lteron} returns a failing result for the CI check; otherwise, it returns a passing result.

\subsection{Implementation Details}

A\textsc{lteron} is implemented as a Python package exposing two CLI entry points, \texttt{alteron} for corpus and snapshot operations and \texttt{alteron-ci} for CI execution. The implementation uses spaCy for dependency parsing and POS tagging, NLTK for real-word collision checks in typo generation, and YAML configuration for MR metadata and CI profiles. All generated artifacts are stored as plain files, so the tool does not depend on any external database and its outputs remain easy to inspect.

Each MR specifies how follow-up inputs are generated, how transformed pairs are validated, and how the expected metamorphic relation is checked. MR transformations are rule-based and seeded rather than LLM-generated, which makes corpus generation deterministic and reproducible. For example, negation MRs use dependency parses to insert negation at controlled syntactic positions, gender-substitution MRs use a constrained lexicon to avoid partial or agreement-breaking swaps, and keyboard-typo MRs use QWERTY adjacency with real-word collision checks. Tokenizer-sensitive MRs such as capitalization changes are checked for applicability during corpus generation and skipped when the tokenizer would make the test invalid.

MR metadata is stored in a YAML registry containing task applicability, expected relation, known failure modes, implementation references, and CI handling categories such as \texttt{hard-fail} and \texttt{soft-warning}. These categories allows the same MR library to support different release policies. The behavioral-regression threshold can be customized globally, per CI profile, or per run. The tool is publicly available and designed for reuse and extension through its CLI, MR registry, and framework-agnostic model loader interface.

\section{Tool Usage}

A\textsc{lteron} is available from the project repository and can be used through its command-line interface.
\vspace{-1mm}
\subsection{Installation}

A\textsc{lteron} requires Python 3.10+ together with the dependencies listed in the repository, including spaCy, the \texttt{en\_core\_web\_sm} model, and the NLTK \texttt{words} corpus.

\subsection{Running the Tool}

A\textsc{lteron} exposes two CLI entry points: \texttt{alteron} for corpus and snapshot operations, and \texttt{alteron-ci} for CI execution. A typical workflow consists of three commands.

\textbf{Corpus generation.} To generate a fixed test corpus from labeled source data and a selected set of metamorphic relations, execute \texttt{alteron corpus generate}.

\textbf{Initial snapshot generation.} To generate the initial snapshot from the initial deployed version, execute \texttt{alteron snapshot baseline}.
A\textsc{lteron} loads models through a framework-agnostic \texttt{model-loader} interface that returns the prediction interface expected by the tool; example loaders are provided in the repository.

\textbf{CI run.} The continuous integration check is run with \texttt{alteron-ci}. It uses a selected CI profile, such as \texttt{pr-fast} or \texttt{release-full}, and returns a pass or fail result that can be consumed directly by CI.

\subsection{Reading the Outputs}
\label{reading_outputs}
The main outputs for inspection are the regression report and the CI summary.

\textbf{Regression report.} The regression report contains one entry per MR, with the following fields:
\begin{itemize}
    \item \texttt{mr\_id}: the MR identifier.
    \item \texttt{source\_accuracy\_delta}: the change in source-side accuracy between the baseline and candidate.
    \item \texttt{n\_matched}: the size of the matched subset used for behavioral comparison.
    \item \texttt{matched\_pass\_rate\_delta}: the change in MR pass rate on the matched subset.
    \item \texttt{behavioral\_regression\_flag}: whether the MR crossed the regression threshold.
    \item \texttt{release\_blocked}: whether the MR contributes to a blocking release decision.
\end{itemize}

\textbf{CI summary.} The CI summary records the overall result of the run, including whether a blocking regression was detected and the final exit status returned by \texttt{alteron-ci}.

The artifact directory also contains the fixed test corpus, the behavioral snapshots, and the rejection logs for further inspection when needed.

\section{Experimental Setup}
\label{experiment}
We evaluate four BERT-family model versions namely \texttt{v1\_base}, \texttt{v2\_retrain}, \texttt{v3\_distilled}, and \texttt{v4\_quantized}. BERT-family encoders were chosen because they remain a standard architecture for supervised NLP classification~\cite{devlin2019bert}. All models are cased encoders, enabling capitalization-sensitive MRs. These versions correspond to three common model update operations, namely retraining on the original task data, knowledge distillation~\cite{sanh2019distilbert}, and post-training 8-bit quantization~\cite{yao2022zeroquant}, which reflect practical updates used to improve task performance or reduce model size and deployment cost.

The evaluation spans nine dataset profiles across three task families. For sentiment analysis, we use SST-2~\cite{wang2018glue} and IMDb~\cite{maas2011imdb}. For natural language inference, we use SNLI~\cite{bowman2015snli} and MultiNLI~\cite{williams2018multinli}. For robustness evaluation, we apply generic MRs to SST-2, IMDb, SNLI, MultiNLI, and AG News~\cite{zhang2015character}. Task-specific profiles are evaluated on their respective datasets, while generic profiles are applied across all datasets to measure robustness under lexical and syntactic perturbations across classification tasks. Overall, the setup includes 9 dataset profiles, 10 MRs, 4 model versions, and 3 model transitions.

\section{Evaluation}
We performed a preliminary system evaluation of \textsc{Alteron} based on the experimental setup in Section \ref{experiment} to assess its ability to identify behavioral regression. Additionally, we conducted an exploratory user study to assess the usability and usefulness of \textsc{Alteron}.

\subsection{System Evaluation of Alteron}
We evaluated A\textsc{lteron} on NLP model updates to assess whether it can identify behavioral regressions that are not reflected in standard benchmark metrics. For each dataset, we built a fixed test set by applying MRs to labeled examples and keeping only valid transformed pairs that pass automated checks. We also logged rejected cases for later inspection. The final test set is stored with a hash and manifest to ensure integrity and reproducibility, and it is reused across all model versions to ensure fair comparison.

We then evaluated all four model versions on this fixed test set and stored their outputs as snapshots. Each snapshot includes predictions for both original and transformed inputs, confidence scores, and whether each MR is satisfied.

Finally, we compare three sequential model updates: \texttt{v1\_base -> v2\_retrain}, \texttt{v2\_retrain -> v3\_distilled}, and \texttt{v3\_distilled -> v4\_quantized}. For each MR, we compute how often both versions correctly classify the original input and measure changes in MR behavior on the transformed inputs. A behavioral regression is flagged when this change falls below a predefined threshold. The decision to block a release depends on MR severity and the active CI configuration. Table~\ref{tab:pilot-summary} summarizes the results across the nine dataset profiles.

In total, A\textsc{lteron} produced 27 regression reports (9 profiles $\times$ 3 transitions) across 81 MR-level comparisons and flagged 16 behavioral regressions, 11 of which were release-blocking hard failures. As shown in Table~\ref{tab:pilot-summary}, all flagged regressions occurred in the generic robustness profiles, with \texttt{gen\_multinli} contributing the largest number of flags, followed by \texttt{gen\_sst2} and \texttt{gen\_imdb}.
\begin{table}[t]
\caption{Pilot summary by profile.}
\label{tab:pilot-summary}
\centering
\footnotesize
\rowcolors{2}{gray!10}{white}
\begin{tabular}{l l l c c r}
\toprule
\textbf{Profile} & \textbf{Task} & \textbf{MR Family} & \textbf{Flags} & \textbf{Block} & \textbf{$\Delta$ Pass-Rate} \\
\midrule
\texttt{sa\_sst2} & SA & Sentiment & 0 & 0 & $-0.0483$ \\
\texttt{sa\_imdb} & SA & Sentiment & 0 & 0 & $-0.0299$ \\
\texttt{nli\_snli} & NLI & NLI & 0 & 0 & $-0.0207$ \\
\texttt{nli\_multinli} & NLI & NLI & 0 & 0 & $-0.0354$ \\
\texttt{gen\_sst2} & SA & Generic & 4 & 3 & $-0.2596$ \\
\texttt{gen\_imdb} & SA & Generic & 3 & 3 & $-0.1905$ \\
\texttt{gen\_snli} & NLI & Generic & 3 & 1 & $-0.0741$ \\
\texttt{gen\_multinli} & NLI & Generic & 6 & 4 & $-0.2444$ \\
\texttt{gen\_agnews} & Topic & Generic & 0 & 0 & $-0.0313$ \\
\bottomrule
\end{tabular}
\end{table}

A representative case appears in the \texttt{gen\_multinli} profile for \texttt{v1\_base -> v2\_retrain}. On space injection invariance MR, source accuracy increased from 0.4647 to 0.7259, yet the matched MR pass rate dropped from 0.6909 to 0.4466 over 7,513 matched examples, yielding a \texttt{matched\_pass\_rate\_delta} of \texttt{-0.2444} and a blocking CI outcome. This shows that a new version can perform better overall while still failing more often on inputs that should preserve the same decision. This was not an isolated case; similar regressions appeared across the generic robustness settings.

Across the three transitions, the pilot suggests that different update types affect behavioral robustness in different ways. Retraining improved source accuracy in some profiles while reducing robustness to surface perturbations. Distillation produced the largest number of flagged regressions, suggesting that compression can preserve top-line performance while weakening stability under transformed inputs~\cite{sanh2019distilbert}. By contrast, quantization produced only one flagged regression in this pilot, suggesting a smaller behavioral impact~\cite{yao2022zeroquant}. This matters in practice because these transitions are common production updates, and gains in accuracy, size, or efficiency can still be accompanied by behavioral regressions across model versions.

\subsection{Exploratory User Study}
We conducted a controlled user study to evaluate how participants interpret behavioral regression results produced by \textsc{Alteron} in a CI setting. The study included eight participants, comprising three ML engineers, one DevOps engineer, and four academic researchers. Participants were given a pre-run model comparison scenario consisting of a fixed test corpus, a stored snapshot of a baseline model, and a candidate model under evaluation. \textsc{Alteron} compares these models using MRs, which test whether the candidate model preserves consistent behavior under controlled input transformations. Participants then inspect the outputs of a CI pipeline, which produces two artifacts: a CI summary and a regression report as mentioned in Section \ref{reading_outputs}. Participants use these outputs to identify which MRs caused regressions, understand how model behavior changed compared to the baseline, and judge whether the changes are severe enough to block deployment.

To evaluate usability and perceived usefulness, participants completed a post-study questionnaire using a 5-point Likert scale ranging from 1 (Strongly disagree) to 5 (Strongly agree). The questionnaire assessed clarity of the system purpose, ease of following instructions, ease of setup, navigability of generated artifacts, interpretability of results, and perceived usefulness of \textsc{Alteron} in CI/CD workflows.

Results show that all eight participants completed the study run. Seven participants correctly identified the baseline and candidate model versions as well as the overall CI outcome, and seven also correctly identified capitalization invariance MR as the blocking MR. Most participants relied primarily on the regression report to determine the source of the failure, and it was consistently rated as the most useful artifact for understanding the CI result. In the usability and usefulness evaluation, six out of eight participants reported that they would be able to use \textsc{Alteron} again with similar guidance, and six participants agreed that the system would be useful in a model maintenance or CI/CD workflow.

\section{Related Work}

Traditional NLP evaluation relies on labeled benchmarks and aggregate metrics such as accuracy. Prior work on behavioral testing has shown that such evaluation can miss important robustness and linguistic failures~\cite{ribeiro2020checklist}. Recent tools show that metamorphic testing can be automated for ML systems. Metamorphic testing has also been used to examine the internal consistency and linguistic properties of deep NLP models~\cite{manino2022systematicity} and to support automated behavioral testing in machine translation~\cite{ferrando2023automating}. LLMORPH automates MR-based testing for large language models across NLP tasks~\cite{cho2025llmorph}, while MDPMORPH applies automated metamorphic testing to deep reinforcement learning agents~\cite{li2025mdpmorph}.

These approaches primarily focus on testing or analyzing a model in isolation. They do not address version-to-version comparison in continuous integration. Alteron is designed for this setting.

\section{Conclusion}

Alteron addresses the problem of behavioral regression in evolving NLP classifiers by treating metamorphic testing as a version-to-version check in continuous integration workflows. It combines a fixed test corpus, per-version snapshots, and matched-subset regression differencing to detect changes that aggregate benchmark metrics alone may miss. In a pilot evaluation spanning nine dataset profiles, four model versions, and three update transitions, Alteron identified release-relevant regressions that were concentrated in generic robustness settings and were not apparent from source-side accuracy alone. These results suggest that common model updates such as retraining, distillation, and quantization can preserve overall task performance while still introducing undesirable behavioral change. Future work includes expanding the MR library, broadening evaluation to more NLP tasks and model families, and refining the tool's usability in larger-scale continuous integration settings.

\end{document}